 \documentclass[preprint,review,12pt]{elsarticle}

\usepackage{amssymb}

\journal{Physics Letters A}

\newcommand{\av}[1]{\ensuremath{\langle#1\rangle}}

\begin{document}

\begin{frontmatter}



\title{Closed form solution for the self-resonances
 in a short  Josephson junction}





\author[1]{S. De Nicola}
\author[2]{ M. Adamo}
\author[2]{E. Sarnelli}
\author[2]{C. Nappi\corref{cor1}}
\ead{c.nappi@cib.na.cnr.it}
\cortext[cor1]{Corresponding author}
\address[1]{CNR Istituto Nazionale di Ottica \\I-80078, Pozzuoli, Napoli, Italy}
\address[2]{CNR Istituto di Cibernetica ''E. Caianiello'' \\
I-80078, Pozzuoli, Napoli, Italy}

\begin{abstract}
We present a closed form solution for the self-resonances in a short
Josephson tunnel junction.  This solution is alternative to the well
known textbook result \cite{Barone,Kulik} based on a series
expansion. Results are derived for the up-to-date case of a $0 -
\pi$ junction.
\end{abstract}


\end{frontmatter}


When a constant voltage $V$ is present across the electrodes of a
Josephson junction, the  current flowing into the junction
oscillates at a frequency $\omega=2 \pi V/\Phi_0$, where
$\Phi_0=2.07\times10^{-15}$ Weber is the flux quantum. On the other
hand, if the major size of the junction $L$ is shorter than the
Josephson penetration depth $\lambda_j$ (short junction limit), the
junction can be viewed as a cavity of length $L$. In the presence of
a uniform magnetic field and at an applied junction voltage $V_n =
\Phi_0 \omega_n /2 \pi$, the oscillation frequency of the Josephson
current matches the n-th harmonic of the junction cavity mode
resulting in the excitation of some of the modes at the frequencies
$\omega_n/2 \pi = n \bar c/2L$ (n = 1,2,3...), where $\bar c$ is the
light velocity of the waves in the resonator. In this case typical
current 'steps', known as 'Fiske steps' \cite{Fiske}, appear in the
$I-V$ characteristic of the junction \cite{Barone}.
\begin{figure}[htbp]
\includegraphics[width=23pc]{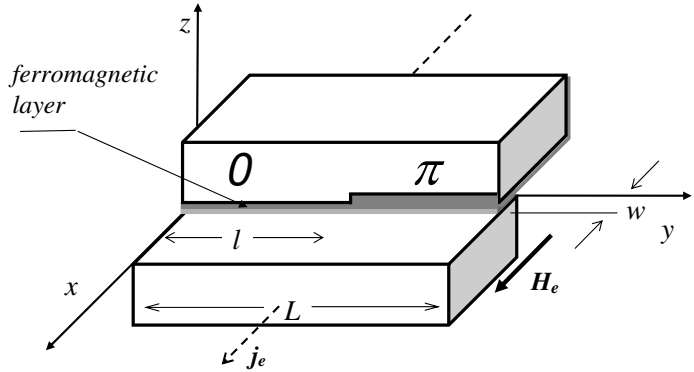}
\caption{\label{sketch} Schematic of the
superconductor-insulator-ferromagnetic metal-superconductor  $0-\pi$
Josephson tunnel junction, along with the coordinate system used in
this work}
\end{figure}

Kulik \cite{Kulik} showed how to calculate, in agreement with the
experiments \cite{Fiske}, the dependence of the amplitude of these
steps as a function of an applied magnetic field in the short
junction limit. In the Kulik's solution, the Josephson phase
difference is expressed in the form of a Fourier series, each term
representing one of the resonances at the voltages $V_n$. Fiske
steps are observed also in long junctions ($L>\lambda_j$), when a
magnetic field is externally applied. In this case however the
junction behaves more like a transmission line than a resonator and
relevant to the underlaying mechanism  is the presence of so called
\emph{fluxons}, particle-like current-field structures periodically
driven inside the junction \cite{Olsen}.

 In this letter we present and discuss an
alternative equivalent approach we have developed for calculating
the Fiske resonances in a short Josephson junction. The method is
based on the development of a closed form solution to the linearized
sine-Gordon equation. Compared to the standard result this method
may present some advantages, in particular if one is interested to a
 description of the phase dynamics given in closed form.
We  apply the method to the relevant case of a $0-\pi$ Josephson
junction, i.e. a junction which has a coupling, changing between $0$
and $\pi$ along the junction length, which implies an anomalous
current-phase relation in the $\pi$-region (see
equation(\ref{eq:J_c})). This physical situation can be realized,
for instance, in superconductor-insulator-ferromagnetic
metal-superconductor tunnel junctions, i.e. junctions in which a
ferromagnetic layer with step-like thickness is inserted, in
addition to an insulating layer \cite{Weides, Weides2} (see figure
\ref{sketch}).

The effect of the presence of few, or many, adjacent $0-\pi$ phase
shifts on the self resonant modes of a Josephson junction has been
considered in the context of  YBCO grain boundary Josephson
junctions \cite{Nappi}. In that reference,  in order to calculate
the contribution of the self-resonances to the current-voltage
characteristics, an extension of the Kulik  expansion was developed.
The results have been successfully used to fit data on Fiske steps
in $0-\pi$ Josephson tunnel junctions \cite{Pfeiffer}. Also
recently, a careful analysis of Fiske modes, based on the Kulik
theory,  has been carried out in
superconductor-insulator-ferromagnetic metal-superconductor to
extract information on the junction quality factor and the relevant
damping mechanisms \cite{Marx}.

 Let us consider a $0-\pi$ junction with two adjacent
 regions $[0,\l]$ and $[\l,L]$, characterized by two different
maximum Josephson current densities, $j_1$ and $j_2$ ($j_1>0$ and
$j_2>0$) and a $\pi$ phase shift in region $[\l,L]$ (See figure
\ref{sketch}). The supercurrent density can be written as
\begin{eqnarray}
\label{eq:J_c} J(y) =  J_c(y)\sin\ \left[\varphi + \pi \theta
(y-l)\right]= \left\{
\begin{array}{rl}
j_1 \sin(\varphi) & \textrm{if $0< y < \l$}\\
j_2 \sin(\varphi+\pi)  & \textrm{if $\l < y < L$}
\end{array} \right.
\label{eq:CPR}
\end{eqnarray}
Assuming a one-dimensional system ($w<<L$), the equation for the
phase difference $\varphi(y,t)$ is expressed by

\begin{eqnarray}
\label{eqn:sine-Gordon} \langle \lambda_j \rangle^2 \frac{\partial
^2\varphi}{\partial y^2}-\frac{1}{\langle \omega_p
\rangle^2}\frac{\partial ^2\varphi}{ \partial t^2}- \frac{1}{{{R C
\langle \omega_p \rangle}^2}}\frac{\partial \varphi}{\partial t}=
 \frac{ J_c(y)}{\langle J_c \rangle}\sin\ \left[\varphi +
\pi \theta (y-l)\right] -\frac{j_{e}}{\langle J_c \rangle}
\end{eqnarray}
where we have introduced the specific normal resistance $R$ and
capacitance $C$ of the junction. Function $\theta(y)$ is the
Heaviside step function, $j_e/\langle{J_c}\rangle$ is the normalized
external bias current density, and  $\langle J_c \rangle
=\left(j_1\l+j_2(L-\l)\right)/L$ is the average maximum Josephson
current density.  In Eq. \ref{eqn:sine-Gordon} the average Josephson
penetration depth is given by $\langle
\lambda_j\rangle=\sqrt{\Phi_0/\left( 2 \pi \mu_0 \langle J_c \rangle
d_{eff}\right)}$ \cite{Nappi}where  the effective depth is $
d_{eff}\approx 2 \lambda_L $ with $ \lambda_L $  the London
penetration length  and we have introduced the average plasma
frequency  $\langle \omega_p\rangle= \sqrt{2 \pi \langle J_c \rangle
/\Phi_0 C}$.

 In the geometry of figure \ref{sketch}, the magnetic field within the
junction is related to the derivative of $\varphi$ through the
relationship
\begin{equation}
\label{eq: field_phiderivative}
 H(y)=-\frac{\Phi_0}{2 \pi \mu_0 d_{eff}} \left(\frac{d\varphi}{dy} \right)
\end{equation}
The boundary conditions appropriate to equation
(\ref{eqn:sine-Gordon}), and to the considered geometry are

\begin{equation}
\label{eqn:bc} \left.\frac{\partial\varphi }{\partial
y}\right|_{0}=\left.\frac{\partial\varphi }{\partial y}\right|_{L}=
-\frac{2 \pi  \mu_0 d_{eff} H_e}{\Phi_0}
\end{equation}

Further important conditions are given by the continuity of
$\varphi$ and  of its  derivative at the point  $y=\l$. The last
condition expresses the continuity of the magnetic field. From now
on, all lengths and time variables will be normalized   to
\av{\lambda_j} and to the inverse average plasma frequency
\av{\omega_p}, respectively, so that equations
(\ref{eqn:sine-Gordon}) and (\ref{eqn:bc}) become respectively
\begin{eqnarray}
  \varphi_{yy}-\varphi_{tt}-\alpha \varphi_t
  = \frac{J_c(y)}{\langle J_c \rangle }\sin\left[ \varphi + \pi \theta(y- l) \right]
  -\frac{j_{e}}{\langle J_c \rangle}
  , \label{eqn:SG_normalized}
\end{eqnarray}

\begin{equation}
  \varphi_y(0,t)=\varphi_y(L,t)=-h_e
  , \label{eqn:bc_normalized}
\end{equation}
where $\alpha = 1/Q=1/\av{\omega_p} R C$ is the dimensionless
damping coefficient, $Q$ the damping quality factor and
$h_e=H_e/\av{\lambda_j}\av{J_c}$ is the normalized external magnetic
field.

Following the Kulik approximation, for solving equation
(\ref{eqn:sine-Gordon}) we write the phase $\varphi$ as a sum of two
terms $ \varphi (y,t)=\varphi_0(y,t)+\varphi_1(y,t)$ where the
unperturbed term is $\varphi_0 = \omega t-h_e y$,
 $\varphi_1\ll \varphi_0$
and $v=\Phi_0/2\pi d \varphi_1/dt$ is the perturbation to the steady
voltage $V$. Here $  \omega= \left( 2\pi V/ \Phi_0
\right)/\av{\omega_p} $ is the normalized Josephson frequency
corresponding to the fixed voltage $V$ applied between the
electrodes of the  junction. We note that, to the zero-$th$  order,
no magnetic field is associated with the $0-\pi$ discontinuity in
the present approximation, as on the left and  right of the point
$y=l$ we have $d \varphi_0/dy=-h_e$, and the only magnetic field
present is the external one. A time dependent magnetic field
perturbation however appears to the 1st order, i.e. $\partial
\varphi_1 /\partial y$.

 If we denote the perturbation $\varphi_1$ as $\varphi_1^L$, when
considered  in the interval $0 \le y \le \l$, and $\varphi_1^R$,
when considered in the interval $\l \le y \le L$, the linear
equations providing $\varphi_1^L$ and $\varphi_1^R$ are respectively

\begin{equation}\label{eq:phip}
      \frac{\partial^2 \varphi_1^L}{\partial y^2}-\frac{\partial^2
      \varphi_1^L}{\partial t^2}-\alpha\frac{\partial \varphi_1^L}{\partial t}=g_1 \sin \left( \omega t
-h_e y\right)
\end{equation}
for $0 \le y \le \l$, and

\begin{equation}\label{eq:phim}
      \frac{\partial^2 \varphi_1^R}{\partial y^2}-\frac{\partial^2 \varphi_1^R}
      {\partial t^2}-\alpha\frac{\partial \varphi_1^R}{\partial t}=g_2 \sin \left( \omega t
-h_e y\right)
\end{equation}
for $\l \le y \le L$. In equations (\ref{eq:phip}) and
(\ref{eq:phim}) we have defined $g_1=j_1/\langle J_c \rangle $,
$g_2=-j_2/\langle J_c \rangle $. The boundary conditions appropriate
to equations (\ref{eq:phip}) and (\ref{eq:phim}) are

\begin{equation}\label{eq:perfect_edges}
      \frac{\partial \varphi_1^L(0,t)}{\partial y}=\frac{\partial \varphi_1^R(L,t)}{\partial
      y}=0
\end{equation}

\begin{equation}\label{eq:phase_continuity}
\varphi_1^L(\l,t)=\varphi_1^R(\l,t)
\end{equation}

\begin{equation}\label{eq:field_continuity}
\frac{\partial \varphi_1^L(\l,t)}{\partial y}=\frac{\partial
\varphi_1^R(\l,t)}{\partial y}
\end{equation}
Equation (\ref{eq:perfect_edges}) is the requirement of perfect
reflectivity of the edges of the junction and continuity of the
phase and of its first derivative at the point  $y=\l$ are
determined by the equations (\ref{eq:phase_continuity}) and
(\ref{eq:field_continuity}), respectively.

 After defining the two
complex functions $u(y)$ and $v(y)$ through the factorization
\begin{equation}\label{eq:phipu}
\varphi_1^L = Re \left[ u(y) \exp(-i\omega t)\right]
\end{equation}
\begin{equation}\label{eq:phimu}
\varphi_1^R = Re \left[ v(y) \exp(-i\omega t)\right]
\end{equation}
and substituting (\ref{eq:phipu}) and (\ref{eq:phimu}) in
(\ref{eq:phip}) and (\ref{eq:phim}), we find that $u$ and $v$
satisfy the two equations
\begin{equation}\label{eq:equ}
\frac{d^2u}{dy^2}+\chi^2 u= i g_1 \exp(i h_e y)
\end{equation}
for $0 \le y \le \l$, and

\begin{equation}\label{eq:eqv}
\frac{d^2v}{dy^2}+\chi^2 v= i g_2 \exp(i h_e y)
\end{equation}
for $\l \le y \le L$, where we have defined $\chi^2=
 \left(\omega^2+i\alpha \omega\right)$. The boundary conditions for the two complex functions  $u(y)$ and
$v(y)$
 are
\begin{equation}\label{eq:bc_usx}
     \frac{du(0)}{dy}=\frac{dv(L)}{dy}=0
\end{equation}
\begin{equation}\label{eq:ueqv}
u(\l)=v(\l)
\end{equation}
\begin{equation}\label{eq:dueqdv}
\frac{du(\l)}{dy}=\frac{dv(\l)}{dy}
\end{equation}
We can write now the general solution to equations (\ref{eq:equ})
and (\ref{eq:eqv}) in the following form
\begin{equation}\label{eq:sol_u}
u(y)=\alpha_u e^{-i \chi y}+\beta_u e^{+i \chi y}+u_p(y),  0 \le y
\le \l
\end{equation}

\begin{equation}\label{eq:sol_v}
v(y)=\alpha_v e^{-i \chi (y)}+ \beta_v e^{+i \chi (y)}+v_p(y), \l
\le y \le L
\end{equation}
where the basic task reduces to determine the four constants
$\alpha_u$, $\beta_u$, $\alpha_v$ , $\beta_v$ from the boundary
conditions (\ref{eq:bc_usx})-(\ref{eq:dueqdv}). In equations
(\ref{eq:sol_u}) and (\ref{eq:sol_v})  we have introduced the
particular solutions  $u_p(y)$ and $v_p(y)$ which, by following
standard methods,  can be promptly written as
\begin{equation}\label{eq:up}
u_p(y)=\frac{i g_1 e^{i h_e y}}{\chi^2-h_e^2}
\end{equation}

\begin{equation}\label{eq:vp}
v_p(y)=\frac{i g_2 e^{i h_e y}}{\chi^2-h_e^2}
\end{equation}
By using the boundary conditions (\ref{eq:bc_usx})-(\ref{eq:dueqdv})
we find the following relationship between the unknown coefficients
$\alpha_u$, $\beta_u$,  $\alpha_v$ and $\beta_v$
\begin{eqnarray}
\label{eqn:system} && i \chi
\left(\alpha_u-\beta_u\right)=\frac{g_1h_e}{\chi^2-h_e^2} \\
\nonumber && i \chi \left(\alpha_v e^{i \chi L}-\beta_v e^{-i \chi
L}\right)=\frac{g_2h_e e^{i h_e L}}{\chi^2-h_e^2}\\ \nonumber &&
\left(\alpha_u-\alpha_v\right)e^{i \chi l}+
\left(\beta_u-\beta_v\right)e^{-i \chi l}=\frac{(g_2-g_1)h_e e^{i
h_e l}}{\chi^2-h_e^2}\\ \nonumber && i \chi
\left(\alpha_u-\alpha_v\right)e^{i \chi l}+i \chi
\left(\beta_u-\beta_v\right)e^{-i \chi l}=\frac{(g_1-g_2)h_e e^{i
h_e
 l}}{\chi^2-h_e^2}
\end{eqnarray}
From equations (\ref{eq:up})-(\ref{eqn:system}) we obtain
$\alpha_u$, $\beta_u$, $\alpha_v$, which can be written as
    \begin{eqnarray}\label{eq:AU}
&& \alpha_u= \frac{1}{2\chi(\sin \chi L)(\chi^2-h_e^2)} \times
\\ \nonumber
&& \left[(g_2-g_1) e^{i h_e l}S(L-l) +h_e (g_1 e^{-i \chi L}-g_2
e^{i h_e L})\right]
\end{eqnarray}

 \begin{eqnarray}\label{eq:BU}
&& \beta_u= \frac{1}{2\chi(\sin \chi L)(\chi^2-h_e^2)} \times
\\ \nonumber
&& \left[(g_2-g_1) e^{i h_e l}S(L-l) +h_e (g_1 e^{i \chi L}-g_2 e^{i
h_e L})\right]
\end{eqnarray}

 \begin{eqnarray}\label{eq:AV}
&& \alpha_v= \frac{1}{2\chi(\sin \chi L)(\chi^2-h_e^2)} \times
\\ \nonumber
&& \left[(g_2-g_1) e^{i (h_e l -\chi l)}S(-l) +h_e (g_1 e^{-i \chi
L}-g_2 e^{i h_e L})\right]
\end{eqnarray}

 \begin{eqnarray}\label{eq:BV}
&& \beta_v= \frac{1}{2\chi(\sin \chi L)(\chi^2-h_e^2)} \times
\\ \nonumber
 &&\left[(g_2-g_1) e^{i (h_e l+\chi l)}S(-l) +h_e (g_1 e^{-i \chi L}-g_2 e^{i h_e L})\right]
\end{eqnarray}
where $S(z)=h_e \cos{(\chi z)}+i \chi \sin{(\chi z)}$.\\ Equations
(\ref{eq:phipu}), (\ref{eq:phimu}), and (\ref{eq:sol_u}),
(\ref{eq:sol_v}), with coefficients (\ref{eq:AU}), (\ref{eq:BU}),
(\ref{eq:AV}) and (\ref{eq:BV}), determine in the present
approximation of 'small $Q$' and short junction \cite{Barone}, the
dynamics of the phase $\varphi_1$ and the magnetic field inside the
junction $\partial \varphi_1/\partial y$ for arbitrary $\omega$ and
$h_e$ values.

 Next, in order to extract a possible dc
term in the current, we have to carry out time and space averages.
That is to say, we have to calculate the quantity
\begin{eqnarray}\label{eq:Jdc}
 &&  J_{dc}(\omega,h_e)= \frac{1}{L}\int_0^L \langle  J_c(y)
   \sin \varphi(y)
   \rangle dy  = \\
   =  \nonumber && \frac{1}{L} \int_0^L \langle J_c(y) \sin (\varphi_0(y)+\varphi_1(y,t)+\pi \theta(\l-L))\rangle dy
 = \\
 \nonumber = && \frac{1}{L}  \left[ j_1 \int_0^{\l} \langle
 \sin (\varphi_0+\varphi_1^L)\rangle
dy - j_2 \int_{\l}^L \langle
 \sin (\varphi_0+\varphi_1^R)\rangle
dy\right]
\end{eqnarray}
\\
Angle brackets indicate time average over the period $T = 2\pi
/\omega$, i.e. if $F(t)$ is an arbitrary function of the time, then
$\langle F(t) \rangle = (1/T) \int_0^T F(t)dt=(1/2 \pi)\int_0^{2
\pi}F(\omega t) d(\omega t)$. Furthermore, since $ \langle
\sin(\varphi_0+\varphi_1)\rangle \approx  \langle  \sin \varphi_0
\rangle +\langle \varphi_1 \cos \varphi_0 \rangle$ and
\begin{eqnarray}\label{eq:average}
&& \langle \sin \varphi_0 \rangle=0 \\ \nonumber && \langle
\varphi_1^L \cos
\varphi_0 \rangle=\frac{1}{2}Re\left[ u e^{-i h_e y}\right]  \\
\nonumber && \langle \varphi_1^R \cos \varphi_0
\rangle=\frac{1}{2}Re\left[ v e^{-i h_e y}\right]
\end{eqnarray}
we obtain for the dc component of the current due to the self-
resonances, the expression

\begin{eqnarray}\label{eq:Jdc}
   \frac{J_{dc}(\omega,h_e)}{\langle J_c \rangle}=  \frac{1}{2L} Re\left[ g_1\int_0^{\l}u e^{-ih_e
y}dy+g_2\int_{\l}^L v e^{-ih_e y}dy\right]
\end{eqnarray}

\begin{figure}[htbp]
\includegraphics[width=20pc]{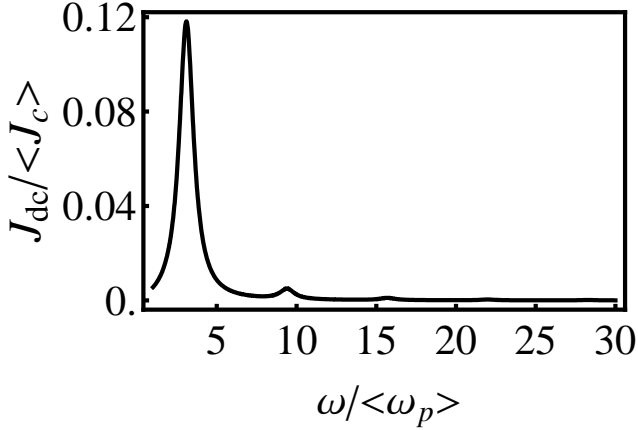}
\caption{\label{fig:IV_zerofield} Current density vs $\omega$ in
zero external magnetic field in a $0-\pi$ Josephson junction;
$\alpha = 1; L = 1; h_e = 0; l = 0.5; g_1 = 1; g_2 = -1;$
self-resonances appear only at the odd positions $\omega = \pi, 3
\pi, 5 \pi $}.
\end{figure}
\begin{figure}[htbp]
\includegraphics[width=20pc]{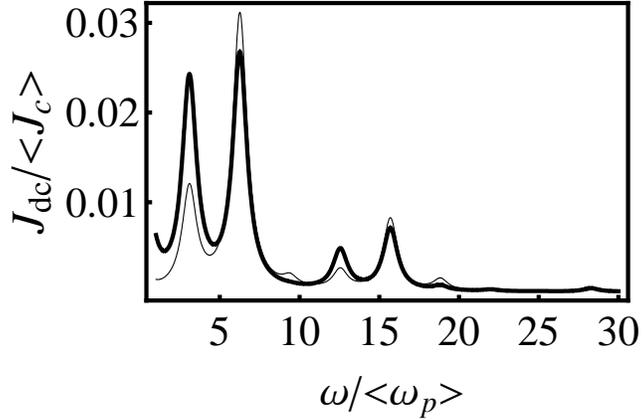}
\caption{\label{fig:comparison} Current density vs $\omega$ in the
presence of an external  magnetic field in a $0-\pi$ Josephson
junction. The two curves refer to two different values of the
magnetic field ($h_e=11$ and $h_e=10$, thin and thick line
respectively); $\alpha = 1; L = 1; l = 0.5; g_1 = 1; g_2 = -1;$}
\end{figure}

The result can be expressed in term of the normalized flux $\phi$
applied to the junction, with the position $h_e L = 2 \pi \phi = 2
\pi \Phi/\Phi_0$, where $\Phi=\mu_0 H_e d_{eff} (L$ and $d_{eff}$
and $L$ are expressed in the usual units in the last expression).
 The
coefficients, equations (\ref{eq:AU}), (\ref{eq:AV}) and
(\ref{eq:BV}), diverge for $\sin{\chi L}=0$ in the limit of
vanishing damping. This gives the resonance frequencies of the
system $\omega_n=n \pi/L$ and, with very good approximation, the
frequencies at which, in the presence of damping, the current
equation (\ref{eq:Jdc}) peaks. The amplitude dependence of the
$n-th$ step on the magnetic field, of fundamental importance for a
comparison with the experiments \cite{Nappi},\cite{Pfeiffer}, can be
calculated by equation (\ref{eq:Jdc}) by setting the value of
$\omega$ at $\omega=\omega_n$.  The same equation can be also used
to probe the 'shape' of the resonances as a function of the
frequency and $Q$ damping factor.
 For the sake of illustration, in figures
\ref{fig:IV_zerofield} and \ref{fig:comparison} we show two cases of
current versus frequency obtained by using equation (\ref{eq:Jdc}).
The first graph refers to the case of zero external field. As can be
seen, odd resonances persist in zero field, even though the
amplitudes of those following the first are vanishingly small, a
phenomenon  typical of the $0-\pi$ junction \cite{Nappi}. The second
one refers to a generic situation of presence of an external
magnetic field and the two curves are calculated at two different
values of the normalized field. The right hand side of each of the
bell shaped peak has negative resistance and, for this reason, has
no relevance for a comparison with the experiments, where usually a
current bias set up is considered. We point out that, in principle,
in the framework of the Kulik theory \cite{Nappi}, to obtain the
same accuracy in the determination of the current density as a
function of frequency or magnetic field, one would have to sum up
the contributions of the entire series representing the current.
 Finally it is worthwhile to stress  that the result
for a $0-0$ tunnel junction, with uniform maximum Josephson current
density, can be easily recovered by the above method. In this case
we have to discuss only the equation
\begin{equation}
\frac{d^2u_0}{dy^2}+\chi^2 u_0= i  \exp(i h_e y)
\end{equation}
with $0\leq y \leq L$, $\varphi_1=Re[u_0(y)exp(-i\omega t)]$ and the
boundary conditions $du_0(0)/dy=du_0(L)/dy=0$. The solution can be
written in the following form
\begin{equation}
u_0(y)=\alpha_{0}e^{-i \chi y} +\beta_{0}e^{i \chi y}+\frac{i e^{i
h_e y}}{\chi^2-h_e^2},
\end{equation}
 only two coefficients  $\alpha_0$ e $\beta_0$  have to be determined. These coefficients are
obtained from the two boundary conditions  since, now, the
continuity condition at $l=0$  is no longer required and they are
given by the following simple expressions
 \begin{eqnarray}\label{eq:A0B0}
 \alpha_0= \frac{1}{2\chi(\sin \chi L)(\chi^2-h_e^2)}
 \left( e^{-i \chi L}- e^{i
h_e L}\right) \\ \nonumber
 \beta_0= \frac{1}{2\chi(\sin \chi L)(\chi^2-h_e^2)}
 \left( e^{i \chi L}- e^{i
h_e L}\right)
\end{eqnarray}
It is easy to verify that these coefficients can be formally
obtained by (\ref{eq:AU})-(\ref{eq:BV}) by letting $l = 0$ and
taking $g_1= g_2=1$. In the same limit, we can also determine from
the general result given by equation (\ref{eq:Jdc}), the explicit
expression of the frequency and field dependance of the current
density, namely

\begin{equation}
J_{dc}(\omega,h_e)=Re \left[ \frac{i h_e^2}{\chi L(\sin \chi
L)(\chi^2-h_e^2)^2} \left(\cos \chi L -\cos h_e L  \right)+ \frac{i
}{2(\chi^2-h_e^2)}\right]
\end{equation}
In conclusion, we have presented and discussed a closed form
solution for the determination of the dynamics of the phase  in a
Josephson junction in the limit of short junction. Within this
framework we have also derived an expression for the dc current
density associated to the Fiske resonances in a  $0-\pi$ Josephson
tunnel junction. This approach can be relevant for improving the
accuracy of data fitting in the determination of the damping
mechanism in $0-\pi$ Josephson junctions or conventional Josephson
tunnel junctions.

\section*{Acknowledgements} This work has been partially supported by
EU STREP project MIDAS, "Macroscopic Interference Devices for Atomic
and Solid State Physics: Quantum Control of Supercurrents".





\bibliographystyle{model1a-num-names}

\begin{thebibliography}{00}


\bibitem{Barone} A. Barone and G. Patern\'o, \emph{Physics and Applications of the Josephson Effect}
 (Wiley, NY 1982)
\bibitem{Kulik}I. Kulik, \emph{JEPT Lett.} \textbf{2}, 84 (1965)
\bibitem{Fiske} M.D. Fiske, \emph{Rev. Mod. Phys. } \textbf{36}(1), 221 (1964)
 \bibitem{Olsen}O. H. Olsen and M. R. Samuelsen, \emph{J. Appl. Phys.} \textbf{52}, 6247
(1981); C. Nappi, M. P. Lisitskiy, G. Rotoli, R. Cristiano, A.
Barone, {\it Phys. Rev. Lett.}, \textbf{93} 187001 (2004)
 \bibitem{Weides}M. Weides, M. Kemmler, H. Kohlstedt, R. Waser, D.
 Koelle, R. Kleiner, E. Goldobin, {\it Phys. Rev. Lett}, \textbf{97}, 247001 (2006)
\bibitem{Weides2} M. Weides, U. Peralagu, H. Kohlstedt, J. Pfeiffer, M. Kemmler,
C. Guerlich, E. Goldobin, D. Koelle, and R. Kleiner, {\it Supercond.
Sci. Technol.} \textbf{23}, 095007 (2010).
\bibitem{Nappi} C. Nappi, E. Sarnelli, M. Adamo, M. A. Navacerrada, {\it Phys. Rev. B}, \textbf{74}, 144504 (2006)
\bibitem{Pfeiffer} J. Pfeiffer, M. Kemmler, D. Koelle, R. Kleiner, E. Goldobin, M. Weides,
A. K. Feofanov, J. Lisenfeld, A. V. Ustinov {\it Phys. Rev. B},
\textbf{77}, 214506 (2008)
\bibitem{Marx} G. Wild, C. Probst, A. Marx, and R. Gross, {\it Eur. Phys. J. B} \textbf{78}, 509–523 (2010)
 \end{thebibliography}



\end{document}